\documentclass[12pt,a4paper]{article}

\usepackage[dvips]{color}
\usepackage{graphicx}
\usepackage{amsmath}
\usepackage{amssymb}
\usepackage{amsfonts}
\usepackage[round,authoryear]{natbib}
\usepackage{setspace}

\begin{document}
\onehalfspacing

\title{Local description of the two-dimensional flow of foam through a contraction}

\author{Benjamin Dollet\thanks{Institut de Physique de Rennes, UMR 6251 CNRS/Universit\'e
de Rennes 1, B\^atiment 11A, Campus Beaulieu, 35042 Rennes Cedex,
France. E-mail: \texttt{benjamin.dollet@univ-rennes1.fr}}}

\maketitle

\begin{abstract}
The 2D flow of a foam confined in a Hele-Shaw cell through a
contraction is investigated. Its rheological features are
quantified using image analysis, with measurements of the elastic
stress, rate of plasticity, and velocity. The behavior of the
velocity strongly differs at the contraction entrance, where the
flow is purely convergent, and at the contraction exit, where a
velocity undershoot and a re-focussing of the streamlines are
unraveled. The yielded region, characterized by a significant rate
of plasticity and a maximal stress amplitude, is concentrated
close to the contraction. These qualitative generic trends do not
vary significantly with the flow rate, bubble area and contraction
geometry, which is characteristic of a robust quasistatic regime.
Using surfactants with a high surface viscoelasticity, a marked
dependence of the elastic stress on the velocity is exhibited. The
results show that the rate of plasticity does not only depend on
the local magnitude of the deformation rate, but also crucially on
the orientation of both elastic stresses and deformation rate. It
is also discussed how the viscous friction controls the departure
from the quasistatic regime.
\end{abstract}

\section{Introduction}

Liquid foams are used in many industrial and domestic
applications, such as ore flotation, enhanced oil recovery or
personal care. Many of these uses take advantage of the rich
mechanical behavior of foams \citep{Hohler2005}: under low applied
strain or stress, they are elastic solids, whereas under high
strain or stress, they undergo plastic flow, resulting of many
elementary plastic events, the so-called T1s \citep{Weaire1999},
characterized by the topological rearrangement of four neighboring
bubbles. Therefore, foams belong to the wide class of the complex
fluids. Since their constitutive items, the bubbles, are of
convenient size (typically $10^{-4}$ to $10^{-2}$~m) for
observation, they are particularly suited to relate a macroscopic
mechanical response, measured for instance by rheometry, to the
microstructural behavior, which is of paramount importance to
inspire or test constitutive rheological models.

Since bubbles are very efficient light scatterers, experiments on
3D foams require techniques such as Diffusive-Wave Spectroscopy
\citep{Durian1991,Gopal1995,Earnshaw1995,Hohler1997,Vera2001} or
X-ray tomography \citep{Lambert2007}. The former technique does
not give precisely the bubble shape, and the latter is up to now
limited by its long acquisition time. To overcome these
difficulties, many experiments have been performed on bubble
monolayers, the so-called quasi-2D foams \citep{Vaz2005}, where
all bubbles can be easily imaged up to high frame rates. Since the
seminal study of \citet{Debregeas2001}, many features of the flow
of quasi-2D foams have been quantified, including elastic
\citep{Asipauskas2003} and plastic aspects
\citep{Dennin2004,Dollet2007}, and viscous stresses
\citep{Katgert2008,Katgert2009}. This has enabled to better
understand simple shear phenomena, like localization or
shear-banding
\citep{Wang2006,Janiaud2006,Katgert2008,Langlois2008}, although
there are still open issues, such as the role of disorder
\citep{Katgert2008,Katgert2009} and that of bubble-bubble
interactions \citep{Denkov2009}. Moreover, bubble monolayers are
usually confined by solid plates, except bubble rafts, and the
viscous friction between bubbles and walls can drastically affect
the flow profiles \citep{Wang2006}. This specificity of quasi-2D
foams is a limitation for direct comparison with 3D foam flows.

Recently, tensorial constitutive models for foams, or more
generally viscoelastoplastic materials
\citep{Saramito2007,Benito2008,Cheddadi2008,Saramito2009}, have
been proposed in order to extend predictions of foam flows beyond
pure shear \citep{Marmottant2008}. They essentially rely on a
local coupling between the local magnitude and orientation of T1s,
the deformation rate, and the elastic strain or stress. Therefore,
to assess the validity of these tensorial models, local
measurements of the elasticity, plasticity and flow profiles of
foams flowing in complex geometries is required. However, such
flows have been less investigated than simple rheometric flows:
some experiments have focused on flows past obstacles either in 3D
\citep{Cantat2006b} or in 2D \citep{Dollet2005,Dollet2006}, and
some studies have reported on foam flows through contractions
\citep{Earnshaw1995,Earnshaw1996,Asipauskas2003,Bertho2006}.
\citet{Earnshaw1995,Earnshaw1996} have shown that the rate of
plastic events is globally proportional to the local rate of
strain, consistently with pure shear configurations
\citep{Gopal1995}, but in the absence of in-situ measurements of
bubble deformation, the role of the elastic stress remains
unclear. Hence, the existing studies have not provided a complete
enough description of the flow to constitute stringent tests for
the models.

To help filling this gap, in this paper, a characterization of the
rheological response of a quasi-2D foam flowing towards, through,
and out a contraction is proposed. We introduce the experimental
setup and the methods of image analysis in Sec.
\ref{Sec:Materials_and_methods}. We then describe in details a
``reference" experiment, quantifying the elastic and plastic
behaviors of the foam, and its flow profile (Sec.
\ref{Sec:reference_experiment}). Several control parameters are
then investigated, such as the flow rate, bubble size, contraction
geometry and interfacial properties of the used surfactants (Sec.
\ref{Sec:control_parameters}). We finally discuss and explain the
complex interplays between elasticity, plasticity and flow, and
between surface and bulk rheologies (Sec. \ref{Sec:discussion}),
which make this configuration suited to shed insight on foam
rheology.

\section{Materials and methods} \label{Sec:Materials_and_methods}

\subsection{Experimental setup} \label{Sec:setup}

\begin{figure}
\begin{center}
\includegraphics[width=9cm]{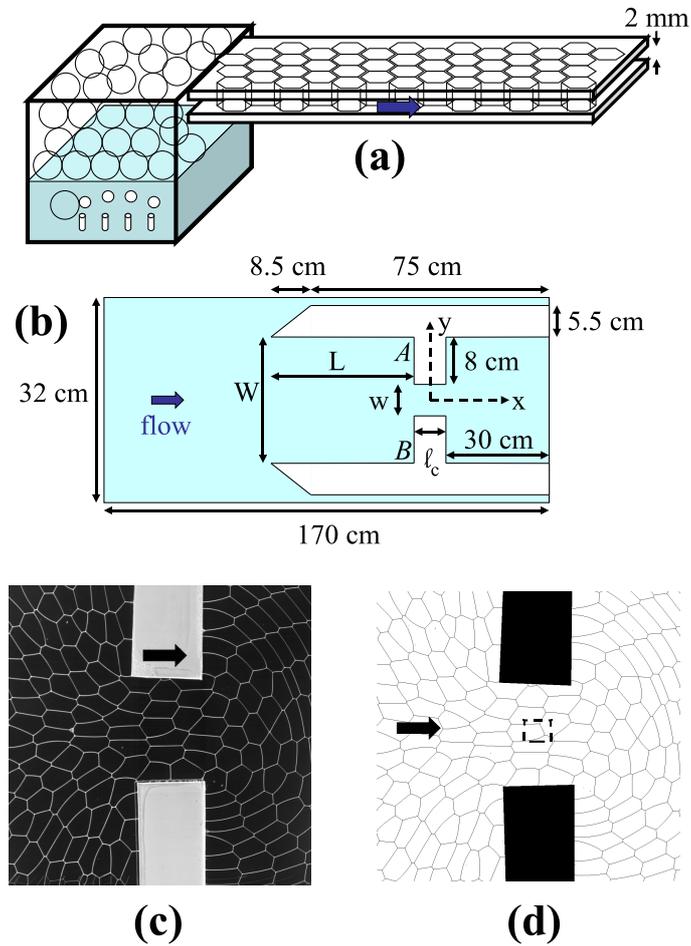}
\caption{\label{Fig:Channel} (a) Side view of the setup (without
the contraction plates). (b) Top view (not to scale) of the foam
channel and the contraction plates. Only the invariant dimensions
have been indicated. The streamwise axis $x$, and the spanwise
axis $y$, originating from the contraction center, are also drawn
in dashed lines. (c) Snapshot of a foam flowing through a
contraction of length 2.0~cm and width 3.2~cm. (d) Skeletonized
version of snapshot (c), used for image analysis. The dashed
rectangle is one of the boxes over which velocity, elastic
stresses and plastic events are counted (see
Sec.~\ref{Sec:image_analysis}). The flow direction is depicted by
the bold arrows in each subfigure.}
\end{center}
\end{figure}

We have used the foam channel fully described in
\citet{Cantat2006}. It is a Hele--Shaw cell, made of two
horizontal glass plates of length 170~cm and width 32~cm,
separated by a gap $h=2$~mm thin enough that the foam is confined
as a bubble monolayer (Fig. \ref{Fig:Channel}a). To make the
contraction, home-made plastic plates were designed. They are
long, narrow plates with a central wider part, which constitutes
the contraction itself (Fig. \ref{Fig:Channel}b). Three pairs were
made, with different contraction lengths $\ell_c$ (2, 5 and 15~cm;
the corresponding lengths $L$ are 35, 33.5 and 28.5~cm). These
plates are inserted in the channel through its open end, such that
the distance between the contraction and the channel exits is
30~cm. This enables to change easily the contraction width $w$.
Its maximal value is 4.4~cm when the plates are in contact with
the channel side walls, and we studied also widths of 3.2, 2.1 and
1.0~cm, for which a small gap remains between the channel side
walls and the plates (Fig. \ref{Fig:Channel}b). The corresponding
aspect ratios $W/w$, with $W$ the channel width upstream and
downstream the contraction, are 4.6, 6.0, 8.6 and 19. The
dimensions and the alignment of the two plates of a given pair are
within 1.5~mm, as can be seen in Fig. \ref{Fig:Channel}c.

The channel is connected upstream to a vertical chamber (Fig.
\ref{Fig:Channel}a) in which a given amount of soap solution is
introduced thanks to a peristaltic pump. Nitrogen is continuously
blown through injectors at the bottom of this chamber, producing
rather monodisperse bubbles (Fig. \ref{Fig:Channel}c) in the flow
rate operating range of the paper (less than 150~ml/min per
injector). The flow rate in each injector is independently
controlled with an electronic flow-rate controller (Brooks). The
resulting foam accumulates on top of the chamber, over a vertical
distance where it drains, then is pushed through the channel. The
transit time through the whole channel is less than 10 minutes in
all experiments; we did not observe significant change of bubble
size during this time, hence coarsening is negligible. The level
of the foam/solution interface in the chamber was kept constant
(within 2~mm) in all experiments to minimize the variations of
liquid fraction. The latter is very low (Fig. \ref{Fig:Channel}c)
and difficult to measure with precision. We can only give a rough
estimate based on the decrease of the level of the foam/solution
interface, which gives the volume of solution evacuated from the
vertical chamber within the flowing foam. Comparing this volume
with the gas flow rate, we get a liquid fraction between 0.2\% and
0.4\% for the experiments presented in this paper.

In order to change the surface rheology, two different solutions
were used. Most of the experiments were done with a solution of
SDS (Sigma-Aldrich; purity $>99.0\%$) dissolved in ultra-pure
water (Millipore) at a concentration of 10~g/l, above the critical
micellar concentration (cmc) of 2.3~g/l. Its surface static and
dynamic properties were measured with a tensiometer (Teclis). We
measured by the rising bubble method a the surface tension $\gamma
= 36.8\pm 0.3$~mN/m. Using the oscillating bubble method, we
measured a total dilatational surface modulus, defined as $E_S =
|\mathrm{d}\gamma/\mathrm{d}\ln A|$ with $A$ the bubble area,
below the noise level, hence lower than 1~mN/m: as expected, pure
SDS interfaces have a negligible viscoelasticity and can be
considered as fully ``mobile" \citep{Denkov2005}. All SDS
solutions were used within a day from fabrication. Conversely, to
study the case highly viscoelastic or ``rigid" interfaces, we used
a mixture of SLES, CAPB and myristic acid (MAc), following the
protocol described in \citet{Golemanov2008}: we prepare a
concentrated solution of 6.6\% wt of SLES and 3.4\% of CAPB in
ultra-pure water, we dissolve 0.4\% wt of MAc by continuously
stirring and heating at 60$^\circ$C for about one hour, and we
dilute 20 times in ultra-pure water. The solution has a surface
tension of $22.0\pm 0.5$~mN/m, and a surface modulus of 216~mN/m
for a frequency of 0.2~Hz and a relative area variation $\delta
S/S_0 = 1.0\%$. Although it fully ensures that the interfaces are
in the ``rigid" limit \citep{Denkov2005}, this value is somewhat
lower than that measured by \citet{Golemanov2008}, which is $E_S =
305$~mN/m. This likely comes from the fact that these authors have
used a piezoelectric control of the bubble oscillations, which was
shown to be of better accuracy to impose sinusoidal area
variations than our syringe-driven control, especially for
interfaces of high surface modulus \citep{Russev2008}. As
\citet{Golemanov2008}, we measured a significant decrease of the
surface modulus with increasing relative area variation (data not
shown). Both solutions have a bulk viscosity equal to that of
water, $\mu = 10^{-3}$~Pa~s.

The contraction region is lit by a circular neon tube of diameter
40~cm, placed just below the channel on a black board. It gives an
isotropic and nearly homogeneous illumination over a diameter of
about 20~cm. Movies of the foam flow are recorded with the
high-speed camera APX-RS (Photron) at a frame rate of 60 or 125
frames per second, with a short shutter speed of 1~ms, so that
even the fastest bubbles (more than 10~cm/s) remain sharp. For all
experiments except the ones with varying width (Sec.
\ref{Sec:geometrical_parameters}) and surfactants (Sec.
\ref{Sec:surfactants}), we have recorded two movies, one upstream
and one downstream the contraction, to record the flow far enough
the contraction. The movies are constituted of 1000~images
constituted by $1024\times 1024$~pixels.

\subsection{Image analysis} \label{Sec:image_analysis}

To extract the relevant rheological information from the movies,
we follow a procedure very similar to that presented in
\citet{Dollet2007}. First, each image of the movie is thresholded
and skeletonized with a custom ImageJ macro. Since the foam is
very dry and therefore the liquid films are thin, the shape of the
bubbles is well preserved (Fig. \ref{Fig:Channel}d), contrary to
wet foams \citep{Dollet2007}. As an important consequence, the
elastic stress, based on the network of bubble edges, can be
precisely estimated. The bubbles touching the contraction plates
(sketched in white in Fig.~\ref{Fig:Channel}b) are less well
preserved, because of the uncertainty on the location of these
plates; notably, their area is often underestimated. Therefore, we
will discard most of the information close to the contraction
plates. Second, the skeletonized movie is analyzed by a custom
Delphi program fully described in \citet{Dollet2007}; based on
individual bubble, edge and vertex tracking, it enables to compute
the velocity, elastic stress and plastic events over a rectangular
mesh of $30\times 30$~boxes covering each image. The area of each
box is comparable with the bubble area (Fig. \ref{Fig:Channel}d),
but the average over the 1000 images of each movie ensures that
the fields are smooth at such a length scale. For the sake of
clarity, the maps of the various fields presented in this paper
are recomputed and displayed at a coarser scale.

The velocity is computed by averaging every bubble displacement
over consecutive images. We will plot it both as a vector field
$\vec{v}$, and as a streamline plot. We also compute the
deformation rate tensor $\bar{\bar{D}} =
(\overline{\overline{\nabla v}} + {^t}\overline{\overline{\nabla
v}})/2$, where $^t$ designs matrix transpose.

To compute the elastic stress \citep{Batchelor1970}, we use the
specific definition valid for 2D foams \citep{Janiaud2005}:
\begin{equation}\label{Eq:elastic_stress_tensor}
\bar{\bar{\sigma}} = \lambda\rho \left\langle
\frac{\vec{\ell}\otimes\vec{\ell}}{\ell} \right\rangle ,
\end{equation}
with $\lambda$ the effective line tension (i.e. the pulling force
exerted by each bubble edge), $\rho$ the areal bubble edge
density, $\vec{\ell}$ the notation for the vector joining two
neighboring vertices (the edge curvature is neglected in the above
expression of the elastic stress), and $\otimes$ the tensorial
product: in components $\alpha,\beta = x,y$, Eq.
(\ref{Eq:elastic_stress_tensor}) can be rewritten as
$\sigma_{\alpha\beta} = \lambda\rho \langle \ell_\alpha
\ell_\beta/\ell \rangle$. Each bubble edge is constituted of a
thin liquid film separated by two parallel vertical interfaces,
between two Plateau borders in contact with the top and bottom
walls. Since the foam is dry, the Plateau borders are of much
smaller size than the gap, so we take the approximation: $\lambda
\simeq 2\gamma h$. Concerning the edge density, since in average a
bubble has six edges, we take $\rho = 3/A$ with $A$ the average
bubble area. The elastic stress, a purely mechanical notion, is
strongly correlated to the purely geometrical notion of bubble
(elastic) deformation, as was proposed in \citet{Aubouy2003} and
shown in various experimental cases
\citep{Asipauskas2003,Janiaud2005,Marmottant2008}; hence, we may
hereafter interpret some features of the elastic stress in terms
of bubble deformation. By definition, the elastic stress is a
symmetric tensor with positive eigenvalues; hence, it is
represented as an ellipse which major (minor) axis is proportional
to the highest (lowest) eigenvalues and oriented along the
corresponding eigenvectors.

The T1s are tracked as described in \citet{Dollet2007}. For the
four bubbles concerned by a T1, we denote $\vec{r}_d$ the vector
linking the centers of the two bubbles that lose contact, and
$\vec{r}_a$ the vector linking the centers of the two bubbles that
come into contact, and we ascribe this information to the box
where the event takes place. In our routine, the research of
appearing and disappearing contacts are run independently;
therefore, to ensure that it is a relevant way to characterize
T1s, we have checked that the number of disappearing ($N_d$) and
appearing ($N_a$) edges is equal (within 10\%) in each box. We
thus compute the scalar field of the frequency of T1s per unit
time and area:
$$ f_{T1} = \frac{N_a + N_d}{2A_{\mathrm{box}} t_{\mathrm{movie}}}
, $$ where $A_{\mathrm{box}}$ is the area of a box and
$t_{\mathrm{movie}}$ the duration of a movie.

As a limitation of our approach, we cannot measure the pressure
field. Laplace law, which states that the pressure differences
between two neighboring bubbles is proportional to the curvature
of their common edge, cannot be used here: it requires to know the
full 3D geometry of the bubbles, whereas the out-of-plane edge
curvature is lost during our purely 2D image analysis.
Furthermore, Laplace law is strictly valid only at equilibrium,
and its applicability to flow situations is questionable.
Nevertheless, we verified in the reference experiment that the
flow rate across various cross-sections remains constant (within
4\%), which is a signature of an incompressible flow. Since the
trace of the deformation rate tensor, $\vec{\nabla} \cdot
\vec{v}$, is zero for such a flow, the deformation rate is a
symmetric and (almost) traceless tensor, hence it has a positive
and a negative eigenvalue: we represent two orthogonal lines, a
thick (thin) one in the direction of the eigenvector associated to
the positive (negative) eigenvalue, i.e in the direction of local
elongation (compression) rate, as in \citet{Dollet2007}.

\section{Study of a reference experiment} \label{Sec:reference_experiment}

We now study in detail a reference experiment. It is a flow of a
SDS foam, at flow rate 150~ml/min, in a contraction of length 2~cm
and width 3.2~cm. The average bubble area is 39~mm$^2$, and the
polydispersity index, defined as the standard deviation of the
list of individual bubble areas, is 22\%, but 90\% of the bubbles
are within 3\% from the average area, the rest being constituted
mainly of much smaller bubbles. Hence, the foam is rather
monodisperse (Fig. \ref{Fig:Channel}c).

We describe the main features of each field (velocity, Sec.
\ref{Sec:velocity}; elastic stress, Sec. \ref{Sec:elastic_stress}
and T1s, Sec. \ref{Sec:T1}) on the maps, and we report the
evolution of the different components of each field along
different axes: the central axis $y=0$; an off-centered, spanwise
axis located half-way between the central axis and the side walls,
i.e. at $|y|=4.8$~cm, with an average over the axes $y=-4.8;$~cm
and $y=4.8$~cm; a spanwise axis across the flow upstream the
contraction, at $x=-9.6$~cm; and the symmetric spanwise axis
downstream the contraction, at $x=9.6$~cm. For the two latter
axes, since $y=0$ is an axis of symmetry, the fields will be
plotted for $|y|\geq 0$, with an average over $-y$ and $y$.

\subsection{Velocity} \label{Sec:velocity}

\begin{figure}[t]
\begin{center}
\includegraphics[width=0.8\linewidth]{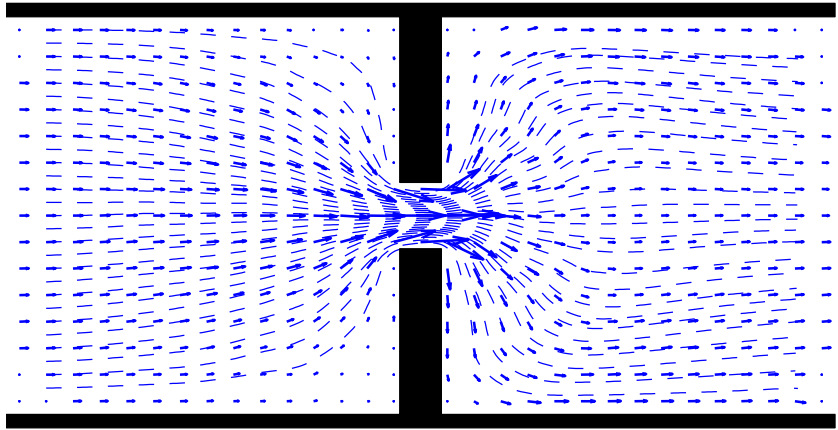}
\caption{\label{Fig:Vitesse_ligne_courant} Velocity field and
streamlines for the reference experiment. The points where the
velocity could not be reliably evaluated are left in blank. The
streamlines starting close to the contraction plates are
interrupted in zones where the velocity could not be reliably
evaluated.}
\end{center}
\end{figure}

\begin{figure}[h]
\begin{center}
\includegraphics[width=0.8\linewidth]{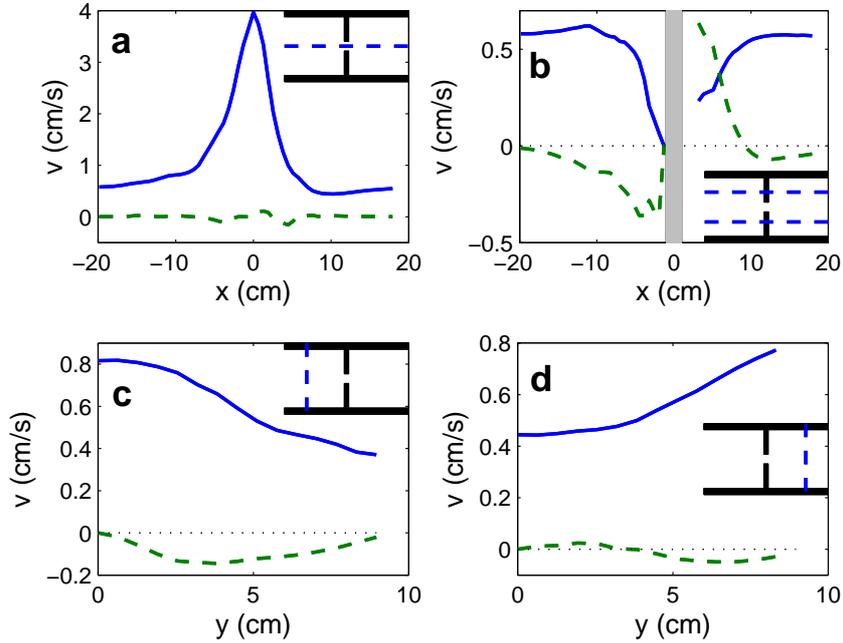}
\caption{\label{Fig:Graphes_vitesse} Plots of the velocity
components $v_x$ (plain curve) and $v_y$ (dashed curve) along (a)
the central axis $y=0$, (b) the off-centered axis $y=4.8$~cm, (c)
upstream the contraction, along the axis $y=-9.6$~cm, and (d)
downstream the contraction, along the axis $y=9.6$~cm.}
\end{center}
\end{figure}

The velocity field and the streamlines are plotted in
Fig.~\ref{Fig:Vitesse_ligne_courant} and the velocity components
along the different axes are plotted in
Fig.~\ref{Fig:Graphes_vitesse}. The foam tends towards a plug flow
far from the contraction, and is faster close to the contraction,
as can be seen from the velocity variation along the central axis
(Fig. \ref{Fig:Graphes_vitesse}a). We do not evidence any dead
zones, nor vortices, in the corners at the entrance side (denoted
by letters A and B in Fig. \ref{Fig:Channel}b); the bubbles keep
flowing towards the contraction even in the corners. This is a
qualitative difference with pure viscoplastic fluids, where dead
zones or vortices are observed in the corners at the entrance side
\citep{Abdali1992,Jay2002}. Moreover, several features show that
the flow displays a strong fore-aft asymmetry with respect to the
contraction center. First, the flow along the channel walls
$y=1$~cm is faster than along the channel walls $y=-1$~cm. Second,
whereas the flow is purely convergent towards the contraction at
the entrance zone, i.e. $v_y$ has always the opposite sign as $y$
(Fig. \ref{Fig:Graphes_vitesse}c), it is not purely divergent at
the exit: Fig. \ref{Fig:Vitesse_ligne_courant} shows that any
streamline passes by a maximum distance from the central axis,
before converging again towards the center. This reversal of the
spanwise velocity component is clearly seen in both on an
off-centered streamwise axis (Fig. \ref{Fig:Graphes_vitesse}b) and
across a spanwise axis in the exit region (Fig.
\ref{Fig:Graphes_vitesse}d). Third, whereas the streamwise
component of velocity along the central axis continuously
increases towards the contraction in the entrance region, it
passes through an undershoot (at $x=10$~cm) in the exit region
(Fig. \ref{Fig:Graphes_vitesse}a); at this point, $v_x$ is 20\%
lower than the plug flow velocity. Furthermore, whereas the
streamwise velocity component decreases from the central axis to
the sides in the entrance region (Fig.
\ref{Fig:Graphes_vitesse}c), it increases in the exit region and
almost doubles from the central axis to the sides (Fig.
\ref{Fig:Graphes_vitesse}d). We finally notice that along the
central axis, $v_y$ slightly deviates from the zero value expected
from symmetry. This is likely due the small geometric
imperfections of the contraction (Sec. \ref{Sec:setup} and Fig.
\ref{Fig:Channel}c).

\subsection{Elastic stress} \label{Sec:elastic_stress}

\begin{figure}[h!]
\begin{center}
\includegraphics[width=0.8\linewidth]{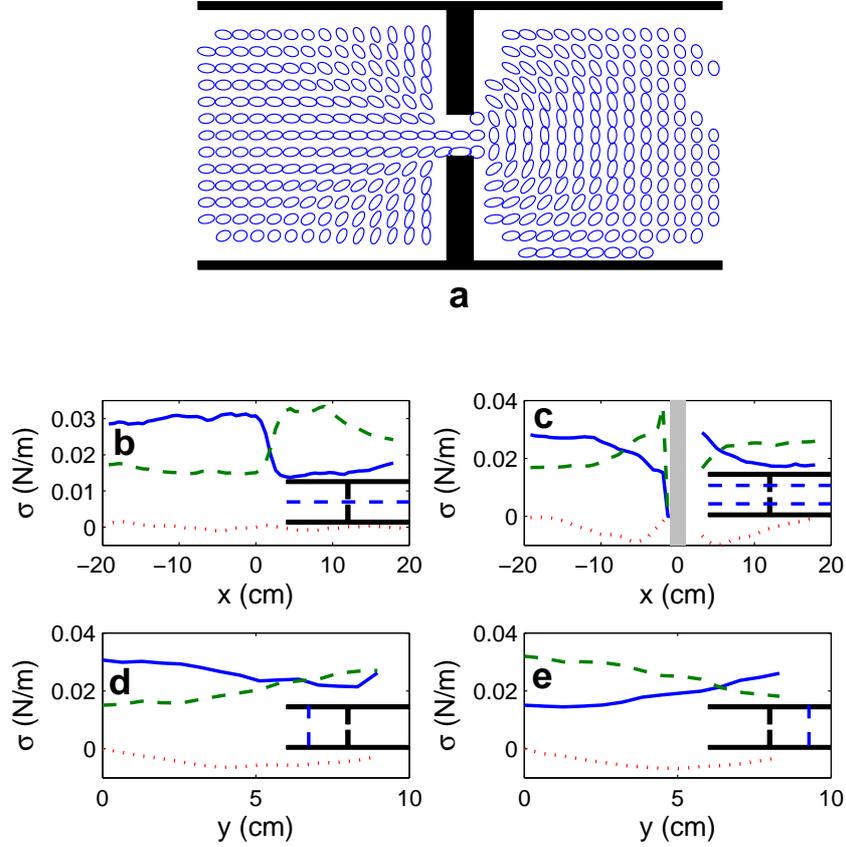}
\caption{\label{Fig:Contrainte} (a) Elastic stress field for the
reference experiment. The major (minor) axis represents the local
direction and magnitude of maximal (minimal) stress, see
Sec.~\ref{Sec:image_analysis}. The points where the elastic stress
could not be reliably evaluated are left in blank. Plots of the
elastic stress components $\sigma_{xx}$ (plain curve),
$\sigma_{xy}$ (dotted curve) and $\sigma_{yy}$ (dashed curve)
along (b) the central axis $y=0$, (c) the off-centered axis
$y=4.8$~cm, (d) upstream the contraction, along the axis
$y=-9.6$~cm, and (e) downstream the contraction, along the axis
$y=9.6$~cm.}
\end{center}
\end{figure}

The map of elastic stress is presented in Fig.
\ref{Fig:Contrainte}a. Not surprisingly, the maximal stress is
directed towards the contraction at the entrance region, because
the bubbles tend to be deformed by the converging flow. The
situation is more complex at the exit region: the bubbles relax
and revert their elastic stress very abruptly just at the
contraction exit, and they are strongly stressed along the
orthoradial direction (from the contraction center) up to a few
centimeters downstream the contraction, before experiencing a
gradual elastic relaxation towards equilibrium, which is not
finished as they are advected away from the observation window. As
a specific feature of the setup, Fig. \ref{Fig:Contrainte}b shows
that the foam is prestressed at the upstream end of the
observation window; indeed, the foam has passed a first
contraction from the main foam channel (of width 32~cm) to the
channel of width $W$ (Fig. \ref{Fig:Channel}) 15~cm only before
the upstream end of the observation window, hence it probably did
not relax completely before arriving in the field of view.

\subsection{T1s} \label{Sec:T1}

\begin{figure}[t!]
\begin{center}
\includegraphics[width=0.8\linewidth]{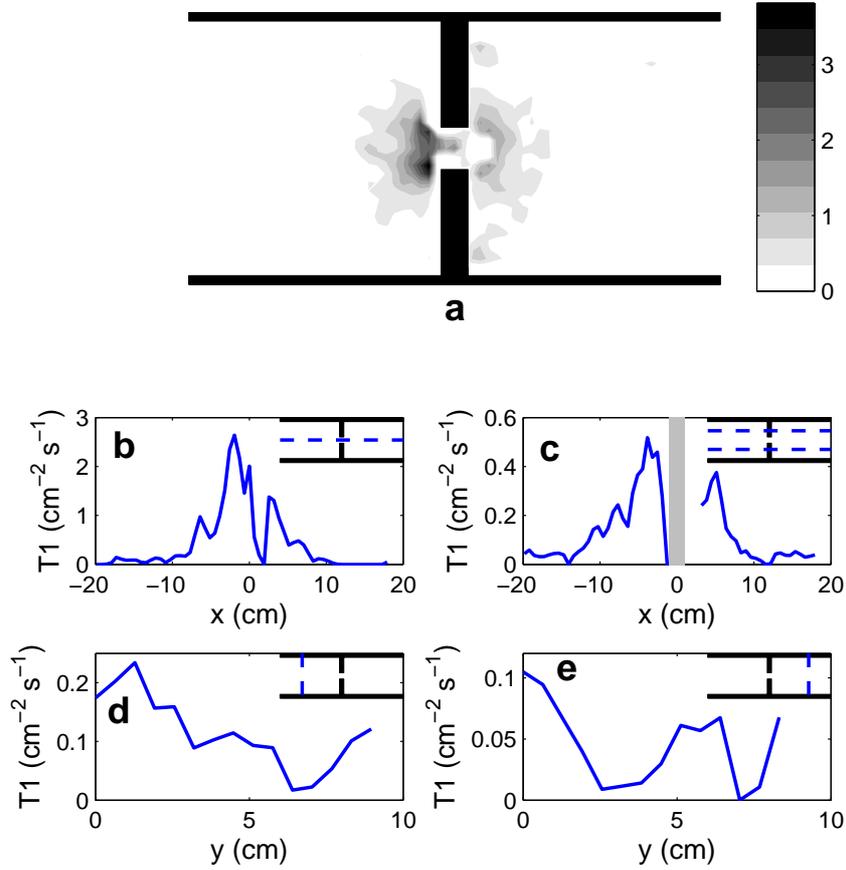}
\caption{\label{Fig:T1} (a) Spatial distribution of the frequency
of T1s, $f_{T1}$, expressed in cm$^{-2}\cdot$s$^{-1}$, for the
reference experiments. Plots of the frequency of T1s along (b) the
central axis $y=0$, (c) the off-centered axis $y=4.8$~cm, (d)
upstream the contraction, along the axis $y=-9.6$~cm, and (e)
downstream the contraction, along the axis $y=9.6$~cm.}
\end{center}
\end{figure}

We now turn to the plastic rearrangements T1s. The spatial
distribution of their frequency is plotted in Fig. \ref{Fig:T1}a.
It shows that plasticity occurs rather close to the contraction,
within a radius of 7~cm. This is confirmed by Figs. \ref{Fig:T1}b,
c, d and e: there are about 5 times less T1s along the
off-centered streamwise axis than along the central axis, and more
than 10 times less across the spanwise axes located 9.6~cm from
the contraction center. As an immediate consequence, the foam
behaves as a viscoelastic medium outside this zone, which confirms
that the features reported farther downstream (re-focussing of the
streamlines towards the central axis, velocity undershoot) are
likely due to the elastic nature of the foam. Fig. \ref{Fig:T1}a
and b show that there are more T1s at the entrance than at the
exit of the contraction; however, this may be a specific result of
the setup due to the prestress of the foam arriving in the
contraction, which facilitates the occurrence of T1s. The
distribution of T1s displays more generic fore-aft asymmetries:
there are significant secondary maxima close to the corners in the
exit region, and not in those of the entrance region; and,
remarkably, there is a small zone of negligible plasticity just at
the exit of the contraction (Fig. \ref{Fig:T1}a and b), which we
will describe further in Sec. \ref{Sec:discussion}.

\section{Influence of control parameters} \label{Sec:control_parameters}

We now study the influence of several control parameters on the
flow of foam through a contraction. Starting from the reference
experiment studied in Sec. \ref{Sec:reference_experiment}, we show
that a change in flow rate and in bubble area does not modify the
velocity and elastic stress, up to a rescaling by the flow rate
and bubble area, respectively (Sec.
\ref{Sec:flow_rate_bubble_area}). We then study the influence of
the length and width of the contraction (Sec.
\ref{Sec:geometrical_parameters}). We finally show how the
physico-chemistry, i.e. the used surfactants, can affect the foam
response in velocity and elastic stress (Sec.
\ref{Sec:surfactants}). A summary of the various experiments, and
their main characteristics: applied flow rate, maximal velocity,
soap solution, mean bubble area, polydispersity, contraction width
$w$ and length $\ell_c$, is displayed in
Tab.~\ref{Tab:summary_exp}.

\begin{table}
\begin{small}
  \centering
  \begin{tabular}{|c|c|c|c|c|c|c|c|}
  \hline
  Sec. & flow rate & maximal & solution & mean bubble & polydispersity
  & $w$ & $\ell_c$ \\
  & (ml/min) & velocity (cm/s) & & area (mm$^2$) & index (\%) & (cm) &
  (cm) \\
  \hline
  3 & 150 & 3.97 & SDS & 39 & 22 & 2.0 & 3.2 \\
  4.1 & 75 & 2.01 & SDS & 28 & 36 & 2.0 & 3.2 \\
  4.1 & $2\times 150$ & 7.38 & SDS & 38 & 25 & 2.0 & 3.2 \\
  4.1 & $2\times 75$ & 3.90 & SDS & 29 & 44 & 2.0 & 3.2 \\
  4.2 & 150 & 7.00 & SDS & 33 & 23 & 2.0 & 1.0 \\
  4.2 & 150 & 5.35 & SDS & 31 & 38 & 2.0 & 2.1 \\
  4.2 & 150 & 3.09 & SDS & 35 & 34 & 2.0 & 4.4 \\
  4.2 & 150 & 3.90 & SDS & 33 & 15 & 5.0 & 3.2 \\
  4.2 & $3\times 150$ & 9.84 & SDS & 34 & 36 & 15.0 & 3.2 \\
  4.3 & 20 & 0.26 & * & 19 & 36 & 2.0 & 3.2 \\
  4.3 & 40 & 0.58 & * & 19 & 36 & 2.0 & 3.2 \\
  4.3 & 70 & 1.02 & * & 19 & 36 & 2.0 & 3.2 \\
  \hline
\end{tabular}
  \caption{Summary of the main characteristics of the experiments presented in this paper.
  The star designs the SLES/CAPB/MAc solution described in Sec. \ref{Sec:setup}.
  The fluid fraction is comprised between 0.2\% and 0.4\% in all experiments (Sec. \ref{Sec:setup}).}
  \label{Tab:summary_exp}
\end{small}
\end{table}

\subsection{Flow rate and bubble area: flow rescaling}
\label{Sec:flow_rate_bubble_area}

To study the influence of the applied flow rate starting from the
reference experiment of Sec. \ref{Sec:reference_experiment}, for
which the gas was injected through a single injector at a flow
rate of 150~ml/min, we used a second injector through which gas is
injected at the same flow rate, independently controlled by a
second flow-rate controller. To study the influence of the bubble
area at the reference flow rate of 150~ml/min, since the flow rate
per injector is the key parameter to tune the bubble size, we blow
gas through the two injectors at 75~ml/min each. We also performed
an experiment with a single injector blowing at 75~ml/min. Note
that we were limited to a narrow range of parameters by the number
of injectors bubbling identically.

To compare these four experiments, we focus on the variations on
the central axis of the relevant components of the fields,
properly rescaled. For the velocity, we take $v_x/v_0$, where
$v_0$ is the entrance velocity, i.e. the velocity averaged over a
cross-stream section at the upstream end of the field of view. For
the 75, 150, $2\times 75$ and $2\times 150$ experiments, we found
respectively entrance velocities of 0.29, 0.58, 0.62 and
1.15~cm/s. For the elastic stress, the relevant components are the
normal components $\sigma_{xx}$ and $\sigma_{yy}$. We will
consider the normal stress difference $\sigma_{xx} - \sigma_{yy}$.
Since the elastic stress scales as a bubble characteristic length,
we rescale the normal stress difference by the trace $\sigma_{xx}
+ \sigma_{yy}$, to compare different bubble areas.

\begin{figure}[t]
\begin{center}
\includegraphics[width=\linewidth]{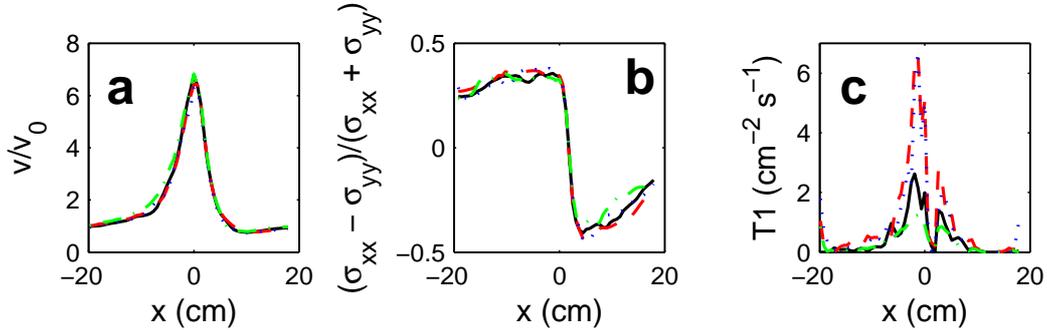}
\caption{\label{Fig:Graphes_A_Q} Comparison between different flow
rates and bubble areas: in terms of number of injectors times the
flow rate per injector (which specifies see bubble area, see text
for details) 150~ml/min (reference experiment, plain curve),
$2\times 150$~ml/min (dashed curve), $2\times 75$~ml/min (dotted
curve), and $75$~ml/min (dash-dotted curve). Plots along the
central axis $x=0$, of (a) the streamwise velocity component
rescaled by the entrance velocity (see text for details),
$v_x/v_0$, (b) the normal stress difference rescaled by the total
elastic stress, $(\sigma_{xx} - \sigma_{yy})/(\sigma_{xx} +
\sigma_{yy})$, and (c) the frequency of T1s.}
\end{center}
\end{figure}

We thus plot along the central axis $v_x/v_0$, $(\sigma_{xx} -
\sigma_{yy})/(\sigma_{xx} + \sigma_{yy})$ and the T1 frequency for
the four experiments, in Fig. \ref{Fig:Graphes_A_Q}. Fig.
\ref{Fig:Graphes_A_Q}a shows that all data for the rescaled
velocity collapse on a single master curve; notably, the velocity
undershoot at $x=10$~cm is a robust observation. Fig.
\ref{Fig:Graphes_A_Q}b shows that the rescaling is good also for
the dimensionless normal stress difference, within slightly
larger, but apparently random, deviations from the main trend.
Since the elastic stress depends on individual bubble sizes, this
may be due to the small variations of polydispersity between the
different runs. However, we consistently observe the same
variations of the dimensionless normal stress difference along the
central axis: a small increase, then a plateau close to the
contraction entrance, followed by a quick reversal just at the
contraction exit followed by an elastic relaxation. Finally, the
T1 frequency shows consistently two peaks, one higher at the
entrance and one smaller at the exit, separated by a narrow zone
with negligible plasticity just at the exit of the contraction
(Fig. \ref{Fig:Graphes_A_Q}c). The T1 frequency for the $2\times
75$ and $2\times 150$ experiments is about the same, and is then
lower for the 150 experiment, and even lower for the 75
experiment. This suggests that the T1 frequency increases as the
velocity increases and the bubble size decreases, and
dimensionally, we could expect that $f_{T1} \propto v/A^{3/2}$.
But if this relation holded, the ratio of the T1 frequency of the
$2\times 150$ and $2\times 75$ experiments would equal 2, whereas
Fig.~\ref{Fig:Graphes_A_Q}c shows that they are almost equal.
Hence, there is no obvious scaling for the T1 frequency.

\subsection{Geometric parameters}
\label{Sec:geometrical_parameters}

We now describe the influence of the geometric parameters of the
contraction itself, i.e. its width and length. First of all,
starting from the reference experiment of Sec.
\ref{Sec:reference_experiment}, we have changed only the width,
and studied four values of it: 1.0, 2.1, 3.2 (reference case) and
4.4~cm. As in the reference experiment, we have used one injector
at 150~ml/min.

\begin{figure}[t]
\begin{center}
\includegraphics[width=\linewidth]{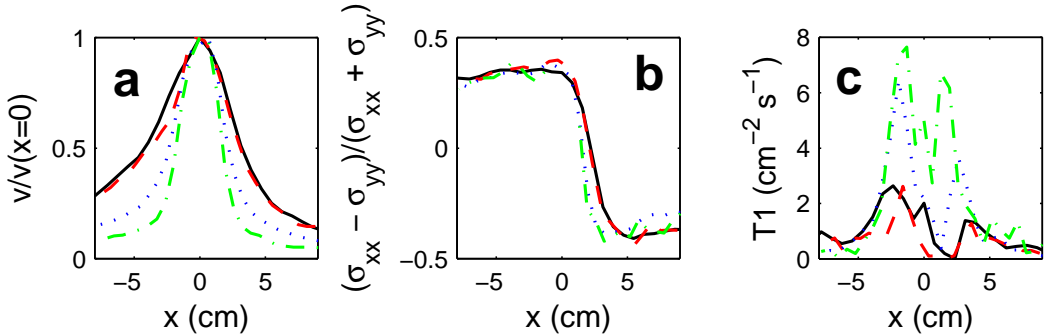}
\caption{\label{Fig:Graphes_largeurs} Comparison between different
contraction widths, at the reference flow rate and bubble area:
1.0~cm (dashed curve), 2.1~cm (dotted curve), 3.2~cm (reference
case, plain curve) and 4.4~cm (dash-dotted curve). Plots along the
central axis $x=0$, of (a) the streamwise velocity component
rescaled by the maximal velocity, $v_x/v(x=0)$, (b) the normal
stress difference rescaled by the total elastic stress,
$(\sigma_{xx} - \sigma_{yy})/(\sigma_{xx} + \sigma_{yy})$, and (c)
the frequency of T1s.}
\end{center}
\end{figure}

As in Sec. \ref{Sec:flow_rate_bubble_area}, a proper rescaling is
useful to compare data on the velocity and elastic stress. A
reference velocity based on the common flow rate of 150~ml/min is
irrelevant, since as the contraction width changes, there is a
varying gap between the contraction plates and the channel side
walls (Fig. \ref{Fig:Channel}) hence a varying ``leakage" flow
rate through this gap. Moreover, for these experiments, we have
recorded only one movie each, centered on the contraction, hence
we resolve the flow on a narrower range up- and downstream the
contraction ($|x|<9$~cm) and we cannot estimate an entrance
velocity in a plug flow regime as in Sec.
\ref{Sec:flow_rate_bubble_area}. Therefore, we take as reference
velocity the maximal velocity (see Tab.~\ref{Tab:summary_exp} for
values), which is reached at the contraction center for these four
experiments. Since the average area varies between the different
runs (Tab.~\ref{Tab:summary_exp}), we also rescale the stress as
in Sec. \ref{Sec:flow_rate_bubble_area}. From the results of last
Section, we expect differences between the four experiments to be
purely due to the width variation. Hence, we plot $v_x/v(x=0)$,
$(\sigma_{xx} - \sigma_{yy})/(\sigma_{xx} + \sigma_{yy})$ and the
T1 frequency along the central axis for the four experiments, in
Fig. \ref{Fig:Graphes_largeurs}.

The modification of the width does not alter the qualitative
trends of the fields. As before, the velocity passes by a maximum
in the contraction (and we miss here the undershoot, too far away
downstream), the stress reverses in the contraction, and the T1
frequency shows two peaks at the entrance and at the exit, the
former being slightly higher than the latter. Basically, the
smaller the width, the more abrupt the velocity and stress
variations across the contraction, and the higher the maxima in
the T1 frequency distribution, which is coherent since the
deformation rate increases.

\begin{figure}[t]
\begin{center}
\includegraphics[width=\linewidth]{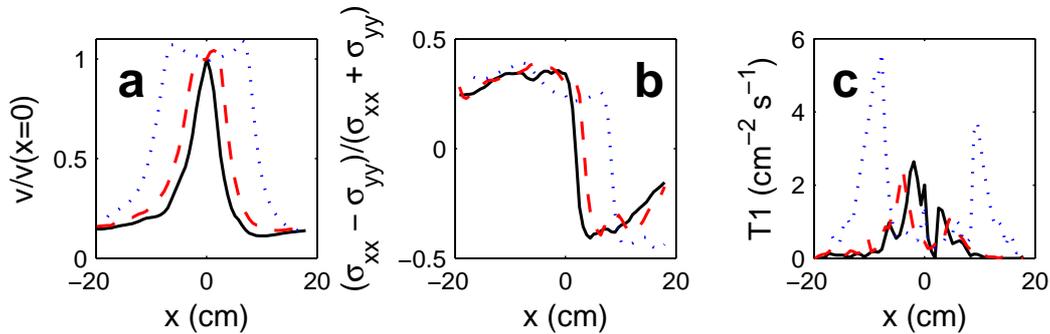}
\caption{\label{Fig:Graphes_longueurs} Comparison between
different contraction lengths, at the reference flow rate, bubble
area and width: 2~cm (reference case, plain curve), 5~cm (dashed
curve) and 15~cm (dotted curve). Plots along the central axis
$x=0$, of (a) the streamwise velocity component rescaled by the
velocity at the middle of the contraction, $v_x/v(x=0)$, (b) the
normal stress difference rescaled by the total elastic stress,
$(\sigma_{xx} - \sigma_{yy})/(\sigma_{xx} + \sigma_{yy})$, and (c)
the frequency of T1s.}
\end{center}
\end{figure}

Second, using different pairs of plates, we have varied the length
of the contraction: from 2~cm (reference case), to 5 and 15~cm.
The width is maintained at the reference value of 3.2~cm. As for
the different widths, we choose to rescale the velocity by
$v(x=0)$, equal to 3.97, 3.90 and 9.84~cm/s respectively for the
lengths 2, 5 and 15~cm, and we plot $v_x/v(x=0)$, $(\sigma_{xx} -
\sigma_{yy})/(\sigma_{xx} + \sigma_{yy})$ and the T1 frequency
along the central axis for the three experiments, in Fig.
\ref{Fig:Graphes_longueurs}.

Fig. \ref{Fig:Graphes_longueurs}a shows a new interesting feature:
for long enough contractions, the velocity is not maximum anymore
at the center of the contraction, as was the case up to now for
the ``short" contraction of 2~cm. Namely, for a contraction length
of 5~cm, the velocity is maximum at $x=1.3$~cm, close to the
contraction exit. For a contraction length of 15~cm, two velocity
overshoots appears, at the entrance ($x=-6.0$~cm) and exit
($x=6.4$~cm) of the contraction; these two overshoots are
respectively 7\% and 9\% higher than $v(x=0)$. Let us notice also
that the velocity inside the contraction does not reach a plateau,
even in the longest contraction, hence its length/width ratio
(equal to 4.7) is not high enough for a plug flow to be fully
established.

Concerning the stress, Fig. \ref{Fig:Graphes_longueurs}b shows
that there is a slow relaxation in the two longest contractions,
but not big enough to come back to equilibrium. For the
contraction length of 15~cm, there is also a small overshoot of
$(\sigma_{xx} - \sigma_{yy})/(\sigma_{xx} + \sigma_{yy})$ before
its fast reversal, which remains a very robust feature of all the
studied flows. The two maxima for the T1s, one stronger at the
entrance and one slightly lower at the exit, remain for the
contraction lengths of 5 and 15~cm; moreover, there is in both
cases a bit of plasticity also within the contraction, associated
to the small stress relaxation.

\subsection{Surfactants} \label{Sec:surfactants}

Up to now, all presented experiments have been performed with a
SDS solution, with a negligible dilatational surface
viscoelasticity. We now study a solution of the surfactant mixture
SLES/CAPB/MAc (see Sec. \ref{Sec:setup}), with a high dilatational
surface viscoelasticity, to quantify the interplay between surface
and bulk rheologies. The geometry of the contraction is that of
the reference experiment: length 2~cm, and width 3.2~cm. Since we
expect the velocity to be the key parameter, we have tried to vary
the applied flow rate at given bubble area. To do so, we have
prepared a foam in the whole channel at a given flow rate of
70~ml/min, and we have changed the flow rate just before
recording; thanks to the big length of the whole foam channel, the
recorded flow is still made of bubbles generated with the flow
rate of 70~ml/min. We have performed three experiments, with
maximal velocity (always reached at the contraction center): 0.26,
0.58 and 1.02~cm/s. We could not reach higher velocities, because
the foam then collapses according to a process which will be
described in subsequent studies. Since we want to track small
structural changes, we have zoomed into the contraction, to get a
better resolution of the bubble shape at the expense of a smaller
field of view (about $7\times 7$~cm$^2$ centered on the
contraction). In order to compare these experiments with the
reference experiment of Sec. \ref{Sec:reference_experiment}, we
plot the streamwise component of velocity rescaled by the maximal
velocity, $v_x/v(x=0)$, and the dimensionless stress difference
$(\sigma_{xx} - \sigma_{yy})/(\sigma_{xx} + \sigma_{yy})$, along
the central axis. In the narrow window of observation considered
here, the statistics of plastic events becomes too poor to be
studied here.

\begin{figure}[htbp]
\begin{center}
\includegraphics[width=0.6\linewidth]{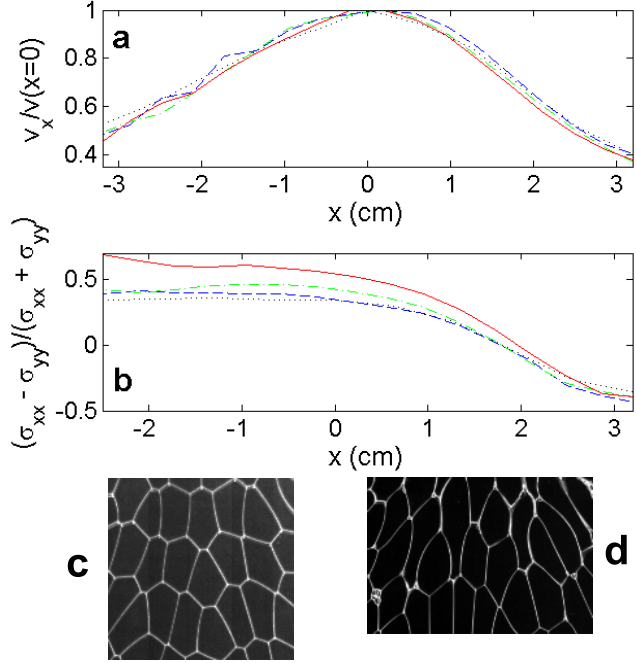}
\caption{\label{Fig:Figure_surfactants} Effect of the surfactant.
Plot of (a) the streamwise velocity component rescaled by the
velocity at the middle of the contraction, $v_x/v(x=0)$, and (b)
the normal stress difference rescaled by the total elastic stress,
$(\sigma_{xx} - \sigma_{yy})/(\sigma_{xx} + \sigma_{yy})$, for
four experiments: the reference experiment with a SDS solution
(dotted curve), and three experiments with the SLES/CAPB/MAc
mixture, with maximal velocity 0.26 (dashed curve), 0.58
(dash-dotted curve) and 1.02~cm/s (plain curve). To illustrate the
change of bubble deformation, snapshots of (c) the reference
experiments and (d) the fastest experiment with the SLES/CAPB/MAc
mixture, taken just at the entrance of the contraction
($-3~\mathrm{cm} < x < -1.5~\mathrm{cm}$), are displayed.}
\end{center}
\end{figure}

Fig. \ref{Fig:Figure_surfactants}a shows that the rescaled
velocity does not vary much between the different experiments, and
that the deviations from the reference experiment remains small.
On the contrary, Fig. \ref{Fig:Figure_surfactants}b shows a strong
and systematic behavior difference between the SDS solution and
the SLES/CAPB/MAc mixture: with the latter, higher values of the
dimensionless stress difference $(\sigma_{xx} -
\sigma_{yy})/(\sigma_{xx} + \sigma_{yy})$ are reached, and the
variation amplitude of this parameter increases with increasing
velocity. As stated in Sec. \ref{Sec:image_analysis}, this is
correlated with a bigger bubble deformation, which is readily seen
by comparison between Figs. \ref{Fig:Figure_surfactants}c and d.

\section{Discussion} \label{Sec:discussion}

\subsection{Interplay between elasticity, plasticity and flow}

In Sec. \ref{Sec:reference_experiment} and
\ref{Sec:control_parameters}, we have described separately the
behavior of the velocity, elastic stress and T1 fields. We now
discuss the coupling between these quantities. First, the
modification of various control parameters, notably flow rate,
bubble area and contraction geometry (Sec.
\ref{Sec:flow_rate_bubble_area} and
\ref{Sec:geometrical_parameters}), has been shown not to change
the \emph{qualitative} features of the flow of foam through a
contraction. The effect of the foam disorder on the flow pattern
remains an open question, as was pointed out by
\citet{Katgert2008}. We have worked with rather monodisperse foams
(Fig. \ref{Fig:Channel}c), but our polydispersity indices are not
negligible, and they vary significantly between various
experiments (Tab.~\ref{Tab:summary_exp}). Hence, the qualitative
trends of the flow of foam through a contraction are not modified
when a moderate amount of disorder is present. Consequently, the
main features of the foam flow are direct consequences of the
interplay of elasticity, plasticity and flow, that we can study on
the reference experiment without much loss of generality.

\begin{figure}[t]
\begin{center}
\includegraphics[width=0.8\linewidth]{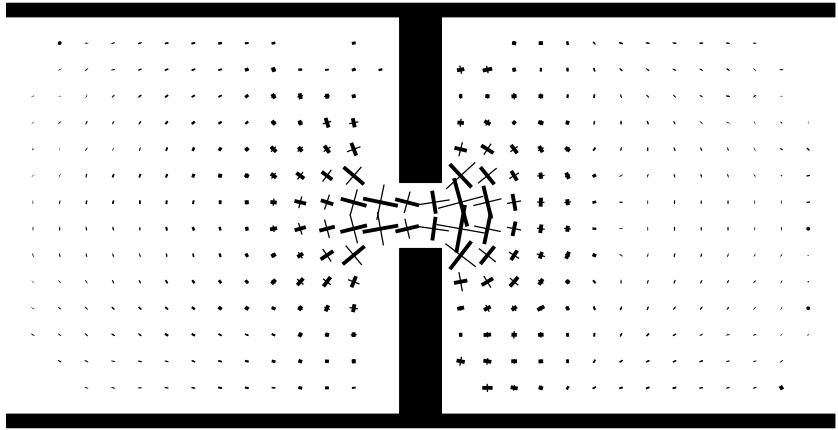}
\caption{\label{Fig:Gradient_vitesse} Deformation rate for the
reference experiment. The thick (thin) lines represent the local
maximal elongation (compression) rate, see Sec.
\ref{Sec:image_analysis}.}
\end{center}
\end{figure}

\begin{figure}[h]
\begin{center}
\includegraphics[width=0.6\linewidth]{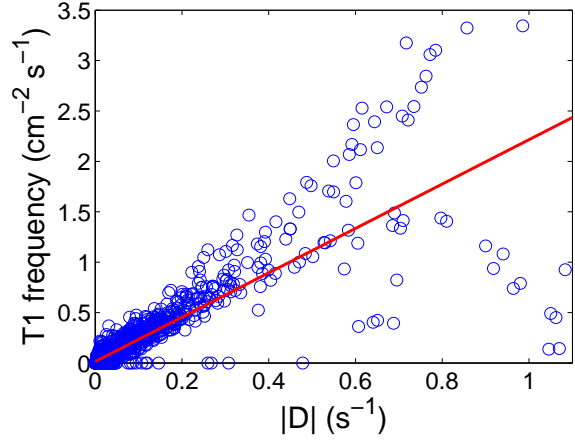}
\caption{\label{Fig:T1(D)} Frequency of T1s \emph{versus}
deformation rate. Each box gives one data point. The straight line
is the best linear fit through all data.}
\end{center}
\end{figure}

Experiments in pure shear \citep{Gopal1995} and contraction flows
\citep{Earnshaw1995} have shown that the rate of plasticity is
globally proportional to the rate of strain. Contrary to these
studies, we have access to the local values of the deformation
rate, which is displayed in Fig.~\ref{Fig:Gradient_vitesse}. From
these local values, we can check whether we recover a
proportionality between the rate of plasticity and the deformation
rate, by plotting for each box the frequency of T1s \emph{versus}
the norm of the deformation rate tensor, $|\bar{\bar{D}}| =
\sqrt{(\bar{\bar{D}} : \bar{\bar{D}})/2} = \sqrt{(D_{xx}^2 +
2D_{xy}^2 + D_{yy}^2)/2}$, in Fig. \ref{Fig:T1(D)}. It shows that
there is a proportionality between the rates of plasticity and
deformation when these are not too large, but that there is a huge
scatter at large rates of plasticity and deformation. Notably,
several points at high deformation rate show very little
plasticity; actually, they correspond to the small zone just at
the contraction exit, where a sharp minimum of plasticity was
already pinpointed in Sec. \ref{Sec:T1}. Hence, the
proportionality between rates of plasticity and deformation is not
valid everywhere.

\begin{figure}[t]
\begin{center}
\includegraphics[width=0.8\linewidth]{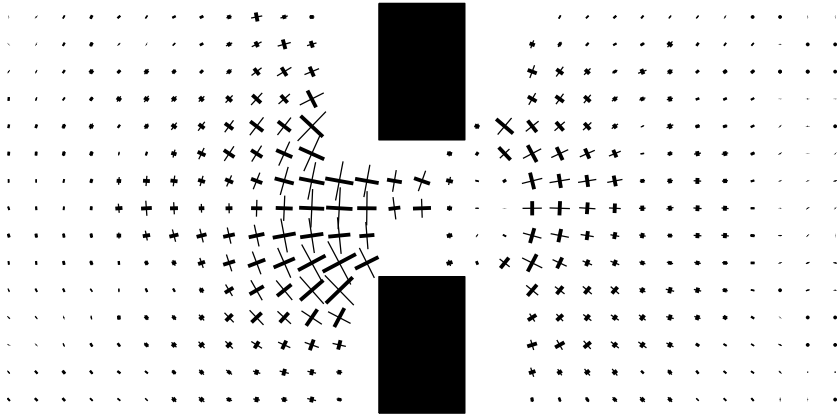}
\caption{\label{Fig:T1_P} Plastic tensor $\bar{\bar{P}}$ for the
reference experiment. The thick (thin) lines represent the
direction of the eigenvector associated to the positive (negative)
eigenvalues, and the line lengths are proportional to the absolute
value of the eigenvalues. The points where the plastic tensor
could not be reliably evaluated are left in blank.}
\end{center}
\end{figure}

We now test a more advanced coupling between elasticity,
plasticity and deformation rate proposed by
\citet{Marmottant2008}. These authors used a plastic tensor
defined as \citep{Graner2008}:
\begin{equation}\label{Eq:plastic_tensor}
\bar{\bar{P}} = \frac{1}{2} f_{T1} \{ (\langle \vec{r}_d \otimes
\vec{r}_d \rangle - \langle \vec{r}_a \otimes \vec{r}_a \rangle)
\cdot \bar{\bar{M}}^{-1} + {^t}[(\langle \vec{r}_d \otimes
\vec{r}_d \rangle - \langle \vec{r}_a \otimes \vec{r}_a \rangle)
\cdot \bar{\bar{M}}^{-1}] \} ,
\end{equation}
where $\bar{\bar{M}} = \langle \vec{r}\otimes\vec{r} \rangle$ is
the texture tensor \citep{Aubouy2003}, defined on the network of
the vectors $\vec{r}$ linking centers of neighboring bubbles. The
plastic tensor encompasses not only the frequency of T1s, but also
there direction; namely, the positive (negative) eigenvalue of
this (almost) traceless tensor is the preferential direction of
bubble separation (attachment). As the deformation rate, we
represent the plastic tensor by a pair of orthogonal lines, the
thick (thin) one in the direction of the positive (negative)
eigenvalue (Fig.~\ref{Fig:T1_P}). As a constitutive relation,
\citet{Marmottant2008} conjectured that the plastic tensor
$\bar{\bar{P}}$, the deviator of the elastic stress tensor,
\begin{equation}\label{Eq:deviator_stress}
\bar{\bar{\sigma}}_d = \bar{\bar{\sigma}} - \frac{1}{2}
(\mathrm{tr}~\bar{\bar{\sigma}}) \bar{\bar{I}} ,
\end{equation}
where $\bar{\bar{I}}$ is the identity tensor, and the deformation
rate $\bar{\bar{D}}$ are related through:
\begin{equation}\label{Eq:model_Marmottant}
\bar{\bar{P}} = \left\{ \begin{array}{l}
  \bar{\bar{0}} ~ \mathrm{if} ~ \bar{\bar{D}} : \bar{\bar{\sigma}}_d \leq 0 \\
  h(\bar{\bar{\sigma}}_d) \dfrac{\bar{\bar{D}} : \bar{\bar{\sigma}}_d}
  {2|\bar{\bar{\sigma}}_d|^2} \bar{\bar{\sigma}}_d
  ~ \mathrm{if} ~ \bar{\bar{D}} : \bar{\bar{\sigma}}_d \geq 0 \\
\end{array} \right. ,
\end{equation}
where $h$, a function varying between 0 and 1, would be the
Heaviside function if the foam showed no plasticity below a yield
stress $\sigma_Y$ and yielded abruptly at that yield stress.
Notice that Eq. (\ref{Eq:model_Marmottant}) slightly differs from
the original version of \citet{Marmottant2008}, where an elastic
strain $\bar{\bar{U}}$ was used instead of $\bar{\bar{\sigma}}$;
but since the deviators $\bar{\bar{\sigma}}_d$ and
$\bar{\bar{U}}_d$ have been shown to be proportional
\citep{Asipauskas2003,Janiaud2005,Marmottant2008}, it is
equivalent to use $\bar{\bar{U}}_d$ or $\bar{\bar{\sigma}}_d$ in
Eq. (\ref{Eq:model_Marmottant}).

The quantity $\bar{\bar{D}} : \bar{\bar{\sigma}}_d$ is
proportional to $\cos 2\theta$, with $0\leq\theta\leq\pi/2$ the
angle between the major axes of elastic stress and deformation
rate; it is positive if $\theta\leq\pi/4$, e.g. if the two axes
are aligned, and negative if $\theta\geq\pi/4$, e.g. if the two
axes are proportional. If the two axes are aligned, $\bar{\bar{D}}
: \bar{\bar{\sigma}}_d = 2|\bar{\bar{D}}||\bar{\bar{\sigma}}_d|$,
hence Eq. (\ref{Eq:model_Marmottant}) predicts that the rate of
plasticity should be proportional to the deformation rate. Since
in most regions of the flow the major axes of elastic stress and
deformation rate are indeed well aligned (Fig.
\ref{Fig:Gradient_vitesse} and \ref{Fig:Contrainte}a), it likely
explains why there is in average a good correlation between the
frequency of T1s and the deformation rate. On the contrary, the
relation (\ref{Eq:model_Marmottant}) implies that no plasticity
should occur if the elastic stress and the deformation rate tend
to act along opposite directions; Fig. \ref{Fig:Gradient_vitesse}
and \ref{Fig:Contrainte}a show that this is the case just at the
contraction exit, and indeed this corresponds to the area with a
marked minimum of plasticity. In this zone, the spanwise
elongation rate serves not to trigger plastic events, but to
reverse the elastic stress from streamwise to spanwise.

\begin{figure}[h]
\begin{center}
\includegraphics[width=0.7\linewidth]{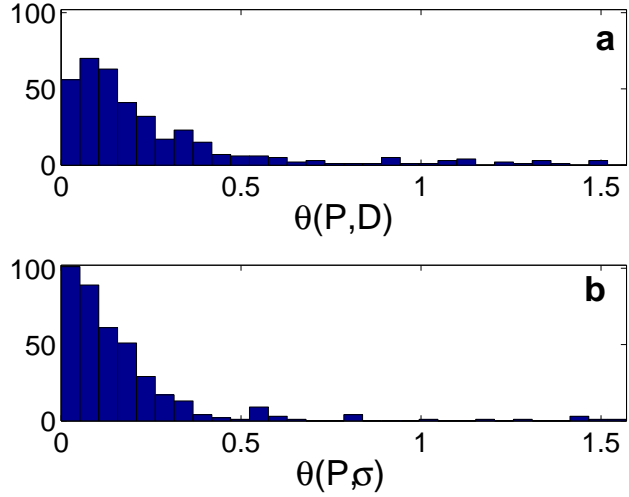}
\caption{\label{Fig:Angles} (a) Histogram of the angle between the
major axes of the plastic tensor and elastic stress,
$\theta(P,\sigma)$. (b) Histogram of the angle between the major
axes of the plastic tensor and deformation rate, $\theta(P,D)$.
Angles are expressed in radians and are defined in the range
$[0,\pi/2]$.}
\end{center}
\end{figure}

There is no model yet for the function $h$ entering Eq.
(\ref{Eq:model_Marmottant}), hence it is difficult to test the
full relation. On the other hand, it predicts that $\bar{\bar{P}}$
should be aligned with $\bar{\bar{\sigma}}_d$ rather than
$\bar{\bar{D}}$, which we test: we compute, for each box, the
angle between the major axes of the plastic tensor and elastic
stress, $\theta(P,\sigma)$, and, for comparison, the corresponding
angle for the plastic tensor and deformation rate, $\theta(P,D)$,
and we plot the histograms of distribution of these two angles in
Fig. \ref{Fig:Angles}. It shows that the plastic tensor is indeed
in average well aligned with the elastic stress than with the
deformation rate; we have $\langle \theta(P,\sigma) \rangle =
0.280~\mathrm{rad} = 16^\circ$, and $\langle \theta(P,D) \rangle =
0.383~\mathrm{rad} = 22^\circ$. This result is a strong indication
that the plasticity rate is not only related to the deformation
rate, but also to the elastic stress. We do not proceed here with
the discussion of the interplay of elasticity, plasticity and flow
and with tests of the models. The next step would be to compare
the full elastic, plastic and velocity fields in our experiments
with simulations and \emph{ab initio} predictions of the flow, but
this is beyond the scope of this paper and will be pursued in
future studies.

At about 7~cm away from the contraction, plasticity vanishes,
hence the foam behaves as a viscoelastic medium. Therefore, the
velocity undershoot (reached in a zone where almost no T1s occur,
see Figs. \ref{Fig:Graphes_vitesse}a and \ref{Fig:T1}b) and the
re-focussing of the streamlines appear as elastic effects. Indeed,
the velocity undershoot is reminiscent of the negative wake
phenomenon \citep{Hassager1979} previously reported also in foams
\citep{Dollet2007}, and attributed to elastic effects; at the
location of the velocity undershoot, the bubbles are very
elongated spanwise, and they then elastically recoil streamwise
(Figs. \ref{Fig:Contrainte}a and b). Similarly, the re-focussing
of streamlines may be attributed to an elastic recoil of the
bubbles.

On the other hand, the velocity overshoots observed at the
entrance and exit of long enough contractions (Fig.
\ref{Fig:Graphes_longueurs}a) appear in yielded regions with a
strong plasticity rate (Fig. \ref{Fig:Graphes_longueurs}c), hence
they do not originate purely from foam elasticity. This is
confirmed by the recent first observation of a velocity overshoot
at the entrance of a contraction for a viscoplastic fluid, by
\citet{Rabideau}. There does not seem to be obvious qualitative
explanations for these velocity overshoots.

\subsection{Interplay between surface and bulk
rheologies}

We have shown in Sec. \ref{Sec:surfactants} that the elastic
stress depends significantly on the velocity, in the case of the
SLES/CAPB/MAc mixture. This interesting observation shows that
when the friction in liquid films, either between neighboring
bubbles or between bubbles and walls, becomes strong enough, a
\emph{dynamic} parameter (here, the local value of the velocity or
of the deformation rate) can influence the structure. This
dynamically-induced deviation from a \emph{quasistatic} foam
structure (such as that fully characterized in Sec.
\ref{Sec:reference_experiment}, which was shown in Sec.
\ref{Sec:flow_rate_bubble_area} indeed not to depend on the flow
rate) has been predicted theoretically by \citet{Kraynik1987} and
observed in other contexts, e.g. by the deviation from the
equilibrium laws governing the angles at which a film meets a wall
\citep{Drenckhan2005}. Actually, this also occurs here: Fig.
\ref{Fig:Figure_surfactants}d shows that some liquid films with
the SLES/CAPB/MAc mixture seem thick, seen from above. This
apparent increased thickness is nothing but the larger horizontal
section of the films that exhibit a strong curvature \emph{across}
the gap between the two plates; this is also related to deviations
from the equilibrium laws, because a liquid film between two
bubbles at equilibrium is straight and perpendicular to the walls.

It would be interesting then to compare our experimental snapshots
with shape predictions from the viscous froth model
\citep{Kern2004}, which is already used to study the morphology of
single films and bubbles \citep{Grassia2008,Cox2009} under the
effect of friction acting on Plateau borders, where a liquid film
meets a wall. However, we have here two sources of friction:
between bubbles and walls, or between neighboring bubbles. When
the latter dominates, which is usually the case in 3D foams, it
has been observed that the yield strain, i.e. the applied strain
at the onset of plasticity, is an increasing function of the
applied strain \emph{rate} \citep{Rouyer2003}. This was indeed
interpreted as a consequence of the internal viscous stresses.
Work is in progress to study the influence of one or the other on
dynamical modification of the foam structure, in simpler
geometries.

Finally, let us stress that such a variation of the elastic stress
is a priori possible also for the SDS solution. However, its
surface viscoelasticity is two orders of magnitude lower than that
of the SLES/CAPB/MAc mixture, hence deviations from the
quasistatic regime are not expected under velocities of order
1~m/s, which is beyond the range of flow rates that we can produce
with our setup.

\section{Conclusions}

We have experimentally investigated the two-dimensional flow of
foam through a contraction. Thanks to image analysis, we could
extract a rather complete information on the foam behavior, in
terms of elastic stresses, plastic events, and flow. The flow
exhibits a strong fore-aft asymmetry (with respect to the
contraction center) and some striking features, such as a velocity
undershoot at the contraction exit. The qualitative features of
the flow are rather insensitive to control parameters such as the
bubble size and contraction width and length, as well as the flow
rate, provided it is low enough (quasistatic regime). Conversely,
as friction within the liquid films becomes significant, for
instance using surfactants with a strong surface viscoelasticity,
the elastic stresses show a strong dependence on the flow
velocity.

By comparing elastic and plastic local descriptors of the foam and
its flow profile, we have been able to show that there is an
essential coupling between elasticity, plasticity and flow.
Notably, there is not everywhere a proportionality between the
rates of plasticity and deformation, and the relative orientation
of elastic stresses and deformation rate plays a fundamental role,
which encourages to further develop and test tensorial
viscoelastoplastic models for foam rheology. As another
perspective of this work, the interesting coupling between surface
and bulk rheologies deserves to be further investigated and
understood, first in simple geometries (plug flow).

\section*{Acknowledgments}

I thank Alain Faisant for the realization of the contraction
plates, Brooks D. Rabideau and Fran\c cois Graner for discussions,
Arnaud Saint-Jalmes and Isabelle Cantat for their careful reading
of the manuscript, and the French ``GdR mousse" for providing a
convenient frame for scientific exchanges and for funding.


\bibliography{"C:/Documents and Settings/bdollet/Desktop/CNRS/Mes publications/Biblio.bib"}

\end{document}